\documentstyle[epsf]{ptptex}

\pubinfo{1996}  
\preprintnumber{
PTPTeX ver.0.7\\ 1996}

\title{ Aging and effective temperatures in the low temperature mode-coupling
equations.   
}

\author{
Leticia F.  {\sc Cugliandolo}\footnote{Also at SPEC, CEA, Saclay.
E-mail address: leticia@lptl.jussieu.fr, leticia@amoco.saclay.cea.fr} 
and Jorge {\sc Kurchan} \footnote{
E-mail address: Jorge.Kurchan@enslapp.ens-lyon.fr}
}

\inst{
$^*$ Laboratoire de Physique Th\'eorique des Liquides, Jussieu, 4 Place Jussieu,
75005 Paris, France 
\\
$^{**}$\'Ecole Normale Sup\'erieure de Lyon, 46, All\'ee d'Italie,
69364 Lyon Cedex 07, France.
}

\recdate{
\today
}

\abst{
The low-temperature generalization  
of the mode-coupling equations corresponds to the dynamics 
of mean-field disordered models in the glassy phase.
The system never achieves equilibrium, preserving the memory
of the time elapsed after the quench throughout its evolution.

A concept of effective temperature can be made quite rigorous
in this context by considering readings of thermometers in different time-scales
and the thermalization of weakly coupled subsystems.
}

\begin{document}

\maketitle

\makeatletter
\if 0\@prtstyle
\def\asp{.3em} \def\bsp{.26em}
\else
\def\asp{.3em} \def\bsp{.3em}
\fi \makeatother

In the past few years progress has been made in the 
analytical  understanding
of glassy dynamics. On the one hand, it was realized that the 
out of equilibrium dynamics of mean-field disordered models was solvable for long times,
and that the solution showed aging phenomena \cite{Cuku} qualitatively close to the ones
of real systems \cite{agingSG,agingPO}.

On the other hand, it became clear that, at least at the mean-field level, the
important distinction  is not between models with and without quenched randomness
\cite{SelfInduced1,SelfInduced2} but rather between different kinds of 
nonequilibrium dynamics.
This became particularly
evident when it was noticed that the mode-coupling equations for supercooled liquids
 \cite{Go}
could be obtained as the dynamical equations for
mean-field disordered models in the high-temperature phase \cite{Kith}.
 By considering the  low-temperature dynamics of these spin-glass models\cite{Cuku}, 
a low-temperature extension of the mode-coupling equations was immediately obtained
\cite{Frhe,Bocukume}.

Indeed, except for the crucial (for structural glasses) question of the behaviour 
near the glass transition, and the related problem of cooling-rate
dependence, the models described above yield a rather realistic picture of
glassy dynamics well below $T_g$.
  
Let us consider $n$ modes $O_1,...,O_n$,
their correlations $C_{ab}(t,t_w) = \langle O_a(t) O_b(t_w) \rangle$
and their responses to perturbations $h_a$ conjugate to $O_a$:
\begin{equation}
\left. 
R_{ab}(t,t_w) \equiv \frac{\delta O_a(t)}{\delta h_b(t_w)}
\right|_{h=0}
\; , \;\;\;\;
 \;\;\;\;\; \chi_{ab}(t,t_w) \equiv \int_{t_w}^{t} dt' \, R_{ab}(t,t')
\; .
\end{equation} 
 
One can the write, in general, Schwinger-Dyson equations \cite{Bocukume}
for their correlations ${\sf C}=(C_{ab})$ and
responses ${\sf R}=(R_{ab})$:
\begin{eqnarray}
{\partial C_{ab}(t,t_w) \over \partial t}
&=&
2 T R_{ab}(t_w,t) 
+ \sum_{c} \left( - \mu_{ac}(t)  C_{cb}(t,t_w) + 
\int_0^{t_w} dt'' D_{ac}(t,t'')  R_{cb}(t_w,t'')
\right)
\nonumber \\
& &
+\sum_{c} \int_0^t dt''  \Sigma_{ac}(t,t'')  C_{cb}(t'',t_w)\;,
\label{eq21a} 
\\
{\partial R_{ab}(t,t_w) \over \partial t}
&=&
 \delta(t-t_w)\delta_{ab} + 
\sum_c \left(
- \mu_{ac}(t)  R_{cb}(t,t_w) + 
\int_{t_w}^t dt'' 
 \Sigma_{ac}(t,t'')  R_{cb}(t'',t_w)
\right)
\;.
\nonumber \\ 
\label{dyson}
\end{eqnarray}

In mean-field models
one can close the Schwinger-Dyson equations into a 
set of dynamical equations involving only 
the correlation and response functions:
\begin{eqnarray}
D_{ab}(t,t')
& = & 
F_{ab} ({\sf C}(t,t')) \;, \;\;\;\;\;\;\;
\Sigma_{ab}(t,t')
=
\sum_{c,d}
 F_{ab,cd} ({\sf C}(t,t')) R_{cd}
\; ,
\\
 F_{ab}({\bf q})
&=&
\frac{\partial F}{\partial q_{ab}}  \;, \;\;\;\;\;\;\; \;\;\;\;\;\;\;\;\;\;\;
 F_{ab,cd}({\bf q})
=\frac{\partial^2 F}{\partial q_{ab}\partial q_{cd}}
\; ,
\label{mctoff1}
\end{eqnarray}
where $F$ is a given function determined by the model. \cite{Bocukume}

If the system equilibrates,  1) the  two-time functions become time-translational
invariant (TTI), 2) there is reciprocity  $C_{ab}(t-t_w)=C_{ba}(t-t_w)$,
and 3) we have the fluctuation-dissipation theorem (FDT):
\begin{equation}
R_{ab}(t-t_w)= \frac{1}{T} \frac{ \partial C_{ab}}{\partial t_w} (t-t_w)
\;,
\;\;\;\;  \;\;\;\;
\chi_{12}(t-t_w) = \frac{1}{T} \; (C_{12}(0)-C_{12}(t-t_w))
\; .
\label{FDT5}
\end{equation}
Putting this information in Eqs.~(\ref{eq21a}) and (\ref{dyson})
 one obtains a single equation for the correlations:
\begin{eqnarray}
{\partial C_{ab}(t-t_w) \over \partial t}
&=&
 - \sum_{c} \mu_{ac} \, C_{cb}(t-t_w) + 
  \frac{1}{T} \sum_{c} \left[
D_{ac}(0)
 \; C_{cb}(t_w-t_w) - D_{ac}^{\infty} C_{cb}^{\infty} 
\right]\nonumber \\
&&+ \frac{1}{T} \sum_{c} \int_{t_w}^t dt''
 \;  D_{ac} (t-t'') \; \frac{ \partial C_{cb}(t''-t_w)} {\partial t''}
\; ,
\label{mcton}
\end{eqnarray}
where $D_{ac}^{\infty} , C_{cb}^{\infty}$ stand for 
$ \lim_{t \rightarrow \infty } C_{ac}(t)$ and 
$D_{ac}^{\infty} \equiv \lim_{t \rightarrow \infty } D_{ac}(t)$, respectively.
Equation (\ref{mcton}) is the usual equilibrium MCT equation for liquids.\cite{Go}
It would be quite difficult to guess (\ref{eq21a}) and (\ref{dyson})
from (\ref{mcton})!

As an example, consider the single-mode case
\begin{equation}
F({\bf q})= q^p 
\; .
\label{psp}
\end{equation}
The mass $\mu(t)$ is chosen so as to
impose normalization at equal times of the autocorrelation $C(t,t)=1$.
With these choices Eq. (\ref{mcton}) is the  simplest Mode-Coupling 
equation proposed by Leutheusser
and Bengtzelius, G\"otze and Sj\"olander. \cite{Begosj,Go}.
The two-time dynamical equations correspond to the low temperature 
dynamics of  a spin-glass model introduced by  Crisanti and Sommers. \cite{Crhoso}

The correlation decay in the high-temperature phase is obtained by solving 
(\ref{mcton}) 
where one sees the familiar $\alpha$ and $\beta$ decay (see Fig. 1).

\setcounter{figure}{0}
\begin{figure}
\input{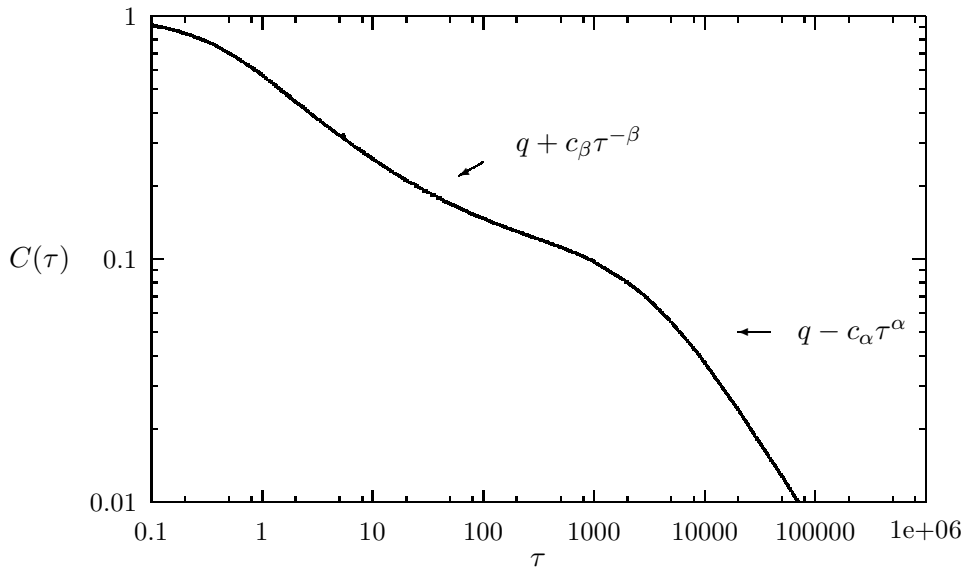}
\caption{ Decay of the auto-correlation function in the high temperature
phase. The plateau develops at $ C=q_{EA}$}
\end{figure}

As one approches the transition temperature from above
 the plateau in $q_{EA}$ becomes larger and larger, but at the same
 time something else happens: 
going back to the original out of equilibrium  equations (\ref{eq21a})
 and (\ref{dyson})
 and solving them  starting from a random  initial condition we discover
 that the time needed to
equilibrate diverges at the transition.
Hence, for temperatures at or below $T_g$ we no longer can assume TTI or FDT, and 
Eq. (\ref{mcton}) becomes irrelevant.

In fact, the two-step process in the correlation decay becomes a waiting-time dependent
 two-step process: the correlations fall below $q_{EA}$ ever more slowly as one
considers an older system (see Fig. 2). Non-trivial $T$-dependent 
exponents \cite{Cule} $\alpha$ and $\beta$ characterize the relaxation around $q_{EA}$.

\begin{figure}
\input{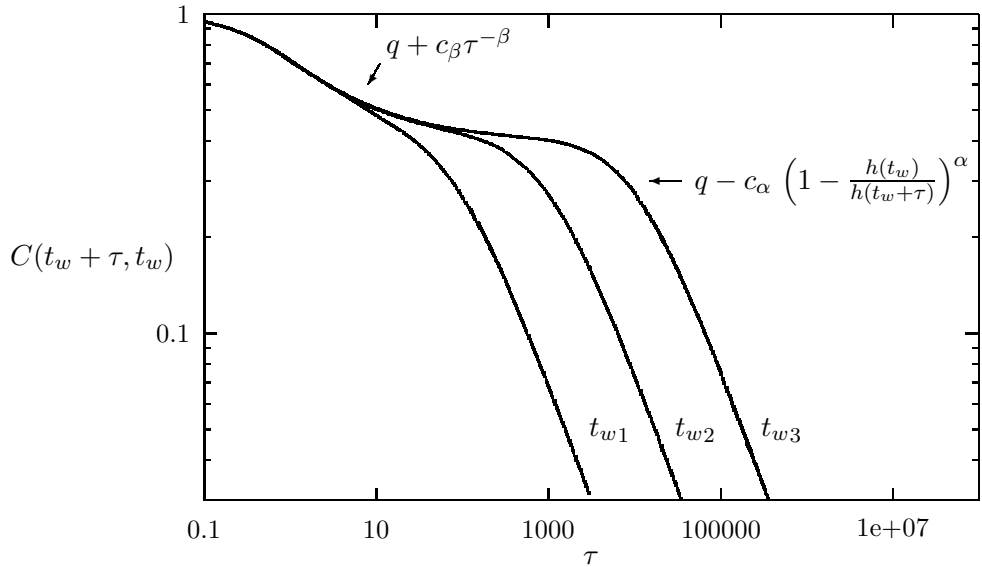}
\caption{Decay of the auto-correlation function in the low 
temperature phase. $C(\tau + t_w, t_w)$ vs. $\tau$ for different waiting times,
$t_w$. The plateau is larger the larger $t_w$. 
}
\end{figure}

We thus see that one of the equilibrium conditions (TTI) is violated: we might 
expect that also the FDT will be violated.
Indeed, this is so: a  parametric plot of  $\chi(t,t_w)$ vs. $C(t,t_w)$
 does not yield
a straight line with gradient $-1/T$ as in equilibrium (cfr. Eq. (\ref{FDT5})),
 but a 
family of curves as in Fig. 3. As one considers larger $t_w$ the curves approach
 {\em two} straight
lines \cite{Cuku}, one with  gradient $-1/T$, corresponding to large and similar times
and another 
 with  gradient $-X/T$ ($X<1$), corresponding to large and widely separated times
(i.e. the aging regime).

\begin{figure}
\centerline{\epsfxsize=10cm
\epsffile{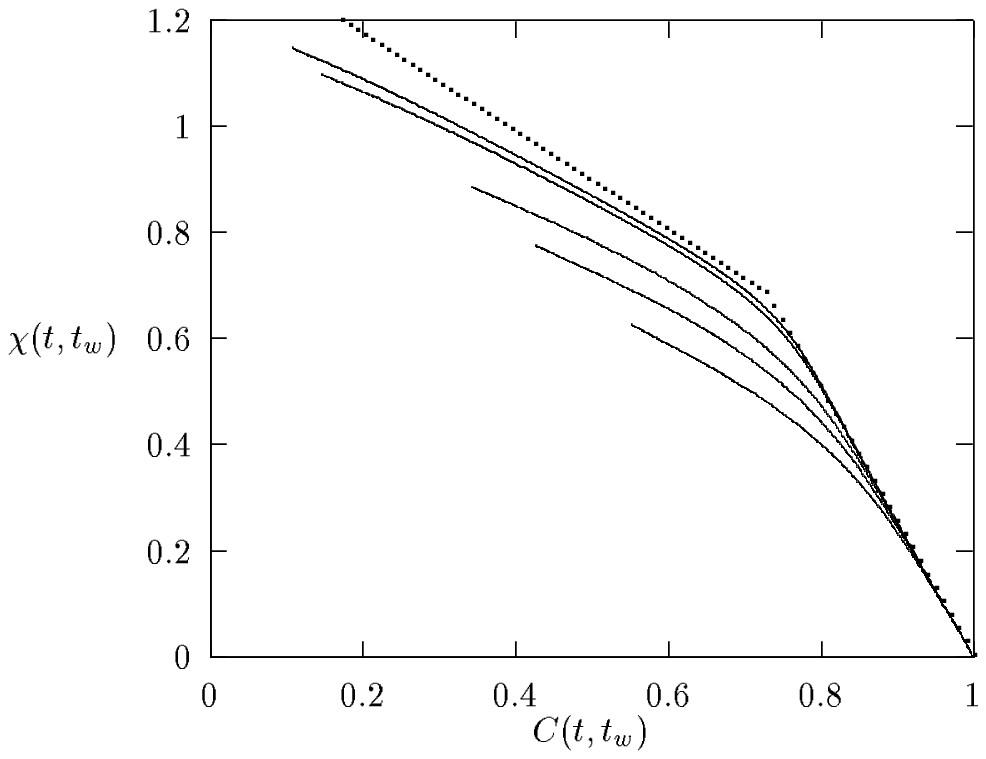}
\label{FIGp3chitfinite}
}
\caption{The susceptibility $\chi(t,t_w)$ vs.\ the auto-correlation 
function $C(t,t_w)$
 at $T<T_g$.  The full curves 
correspond to different total times $t$,  from bottom to top, 
$t=12.5,25,37.5,50,75$.
The dots represent the analytical solution  when $t_w \to \infty$.
}
\end{figure}

The fact that $T/X$ --- the fluctuation-dissipation ratio --- 
might be related to an effective temperature $T_{\mbox{eff}}= T/X$ was noted 
several times, in particular by Hohenberg and Shraiman in the context of 
spatiotemporal chaos and weak turbulence.\cite{turb}

Recently \cite{cukupe},  it was argued that indeed $T_{\mbox{eff}}$ deserves
 the name temperature
in that 1) it is related to the reading of a thermometer that is brought in contact 
with the glass, 2)
it decides the direction of heat-flow and 3) it  is a criterion for 
equilibration.

\begin{figure}
\label{coil}
\centerline{\epsfxsize=10cm
\epsffile{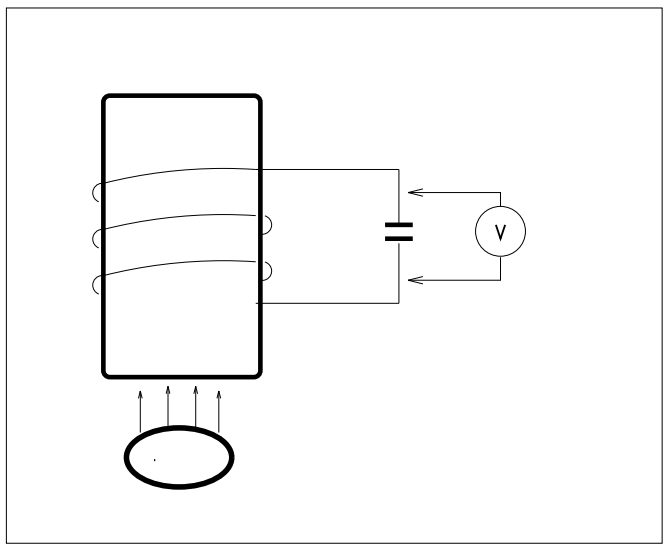}
\label{FIGdibu}
}
\caption{An effective temperature measurement for a magnetic system. The coil
is wound around the sample, which is in contact with the bath. Th `thermometer'
is the $L-C$ circuit.
}
\end{figure}

Let us first consider a very simple thermometer consisting of a harmonic oscillator
that is weakly coupled to an observable $O$ of the glass. For example, 
in Fig. 4 we show how this can be done if $O$ is the magnetization.

In the absence of coupling  $O(t)$ has fluctuations 
with, we assume, zero mean $\langle  O(t) \rangle =0$ and correlations 
$\langle  O(t)O(t_w) \rangle = C_O(t,t_w) = O(N)$. The response  
of the system to a field $h$ conjugate to $O$ is 
$R_O(t,t_w)=\delta \langle O(t)\rangle /\delta h (t_w)|_{h=0}$.

The oscillator takes up energy from
the fluctuations of $O$, and dissipates it through
the response of the system until a  stationary regime is achieved.
If the system is in equilibrium, equipartition of energy implies that
$T=\langle E_{osc} \rangle$ (we have set Boltzmann constant to one).
If we now take $T_O(\omega,t_w)=\langle E_{osc} \rangle$ as 
the natural definition of frequency and time dependent temperature 
of $O$, a simple calculation yields:
\begin{equation}
T_O(\omega,t_w) \equiv  
\langle E_{osc}\rangle_{t_w}=  
 \frac{\omega_o  \tilde C'_O(\omega_o,t_w)}{ 
  \chi_O''(\omega_o,t_w)}
\; ,
\label{effectT}
\end{equation}
where we have used the waiting-time dependent susceptibility and correlations
defined from: 
\begin{eqnarray}
\left[\chi'(\omega,t) + i  \chi''(\omega,t) \right]\;\exp(i \omega t)
 &\equiv& \int_0^t dt' \;
 R(t,t') \exp(i \omega t')
\; , \nonumber \\
\left[ {\tilde C}'(\omega,t)+i {\tilde C}''(\omega,t) \right]
 \; \exp(i \omega t)  &\equiv& \int_0^t dt' \; C(t,t') \exp(i \omega t')  
\; .
\label{ass2} 
\end{eqnarray}

If a system is in equilibrium, FDT holds and $T_O(\omega,t_w)$ is independent
of the observable $O$, the waiting time $t_w$ and the frequency $\omega$, and
is equal to the temperature of the bath.

In the system defined above, we obtain for large frequencies
 $T_O(\omega,t_w)=T$ as $\omega \rightarrow \infty$. If instead we consider
a frequency-waiting time domain such that \linebreak
$C(t_w+\frac{1}{\omega},t_w)<q_{EA}$
(i.e. we are probing the aging scale), then:  $T_O(\omega,t_w)=T/X>T$.

Consider now an experiment
in which we connect the oscillator to an observable $O_1$ and let it
equilibrate at the temperature $T_{O_1}(\omega,t_w)$,
after that we  disconnect it and connect it to another observable $O_2$
and let it equilibrate at the temperature $T_{O_2}(\omega,t_w)$. 
The net result is that an amount of energy $T_{O_1}(\omega,t_w)-
T_{O_2}(\omega,t_w)$ was transferred from the degrees
of freedom associated with $O_1$ to those associated with $O_2$:
the flow goes from high to low temperatures.

This is somewhat like 
touching two points of a glass that has been thermalizing for a long time
 with a copper wire. Even if the glass is still out of equilibrium we would
be surprised if one could obtain an energy flow through the wire this way.
Proposing that this cannot happen is equivalent to saying that for an `old'
glass, different observables should have in the same frequency range the
same temperature.

In order to test this idea, we enlarge the model we have been discussing by
considering two modes, with 
 $F$ given by:
\begin{equation}
F({\bf q})= q_{11}^p + K^2 q_{22}^p
\; .
\label{pspin}
\end{equation}
We impose normalization at equal times of the autocorrelation of both modes:
\begin{equation}
C_{11}(t,t)= C_{22}(t,t)=1
\; .
\end{equation}
through two Lagrange multipliers $\mu_{11}(t)$,  $\mu_{22}(t)$.
We couple both modes through $\mu_{12}=\mu_{21}=\epsilon$
(cfr. Eqs.~(\ref{eq21a}) and (\ref{dyson})).
The constant $K \sim 0.7$ was introduced in  order to break the symmetry
between the modes.
 
Remarkably, it turns out that one can close the equations 
with two ansatze for the long-time
aging behaviour.
In terms of the effective temperatures 
 their meaning is:
\begin{enumerate}
\item {\it Thermalized aging  regime.}
 The effective temperatures associated with the observables $O_1$, $O_2$
 are  equal to each other for frequencies and waiting times
 in the aging regime but different from   the temperature of the bath.
At higher frequencies, their temperatures coincide with the one of the bath:
\begin{eqnarray}
T_1(\omega,t_w)&=&T_2(\omega,t_w)=T\;,\quad
t_w \rightarrow \infty\;, \;\;\;\;\;\;\;\;\;\;\;\;\;\;\;\;\;
C_{ab}>q_{ab}^{EA}\;,\qquad
 {\mbox{quasieq.}} \; ,
\nonumber
\\
T_1(\omega,t_w)&=&T_2(\omega,t_w) \neq T\;,\quad
\omega  \rightarrow 0\;, \;\;t_w \to \infty \;,\quad 
C_{ab}<q_{ab}^{EA}\;,\qquad
\mbox{aging} \; .
\nonumber
\end{eqnarray}

Not surprisingly, in this  case we find that $O_1$ and $O_2$ are
 strongly coupled ({\em also in the aging regime\/}) in the sense that
the cross responses in the aging regime 
\begin{equation}
{R}_{12}(t,t_w)= 
\frac{X}{T} \frac{\partial  {C}_{12}}{\partial t_w} \;
 ,\qquad{R}_{21}(t,t_w)= \frac{X}{T} \frac{\partial  {C}_{21}}{\partial t_w}\;,
\end{equation}
are of the the same order of the self response functions ($X > 0$). 

\item {\it Unthermalized aging  regime.}
 The effective temperatures associated with the observables $O_1$, $O_2$ 
 for combinations of 
frequencies and waiting times corresponding to the aging regime are
neither equal to each other nor to that of the bath, while for higher frequencies
they both coincide with the one of the bath.
\begin{eqnarray}
T_1(\omega,t_w) &=& T_2(\omega,t_w)=T \;,\quad 
t_w \rightarrow \infty\;,\;\;\;\;\;\;\;\;\;\;\;\;\;\;\;\;\;
C_{ab}>C_{ab}^{EA}\;, \;\;\;  
{\mbox{quasieq.}}
\nonumber
\\ 
T_1(\omega,t_w) &\neq& T_2(\omega,t_w) \neq T \;,\quad 
\omega  \rightarrow 0\;, \; t_w \to \infty \;,\quad 
C_{ab}<C_{ab}^{EA}\;,\;\;\; 
\mbox{aging}
\; . 
\nonumber
\end{eqnarray}
In this case, 
$O_1$ and $O_2$ are effectively uncoupled ({\em in the aging regime}):
\begin{equation}
{R}_{12}(t,t_w) = \frac{X_{12}}{T} \frac{\partial  {C}_{12}}{\partial t_w}\;;
\qquad{R}_{21}(t,t_w) = \frac{X_{21}}{T} \frac{\partial  {C}_{21}}{\partial t_w}\;;
\end{equation}
with $ X_{12} \; \rightarrow 0$ and  $ X_{21} \; \rightarrow 0$.
\end{enumerate}

  In Fig. 5 we plot $\chi(C)$
for the  two  uncoupled systems ($\epsilon=0$)  evolving from the initial 
condition $C_{11}(0,0)=C_{22}(0,0)=1$ and $C_{12}(0,0)=C_{21}(0,0)=0$.
We see that  $X_{11} \neq X_{22}$ 
while $X_{12}= X_{21}=0$ (unthermalized case). In Fig. 6
 we consider the same two systems, this time coupled weakly 
($\epsilon = 0.7$) 
Clearly, after a short transient associated to short times, all curves
$\chi_{ij}(C_{ij})$ become  parallel. 
The aging-regime temperatures for the two subsystems 
have  become equal (thermalized case). 

\begin{figure}
\centerline{\epsfxsize=10cm
\epsffile{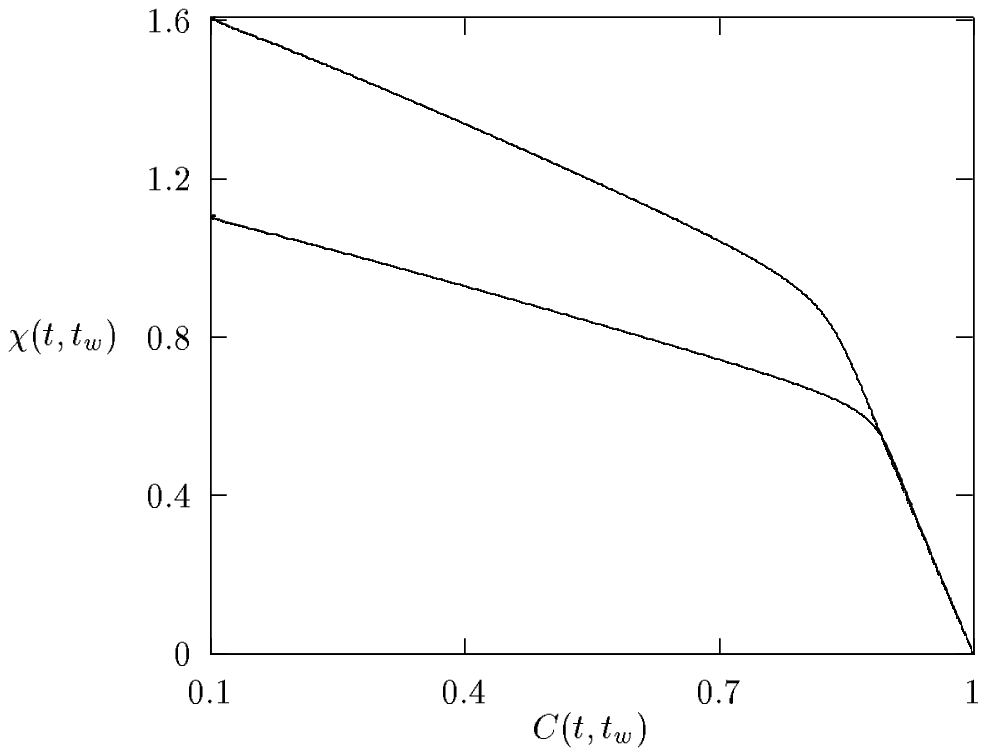}
\label{FIG2syste0}
}
\caption{The susceptibility $\chi_{11}(t,t_w)$ and $\chi_{22}(t,t_w)$ 
vs.\ the corresponding 
auto-correlation functions $C_{11}(t,t_w)$ and $C_{22}(t,t_w)$
for the   {\it uncoupled}, aging systems. The effective temperatures
in the aging regime 
are different.
}
\end{figure}
\begin{figure}
\centerline{\epsfxsize=10cm
\epsffile{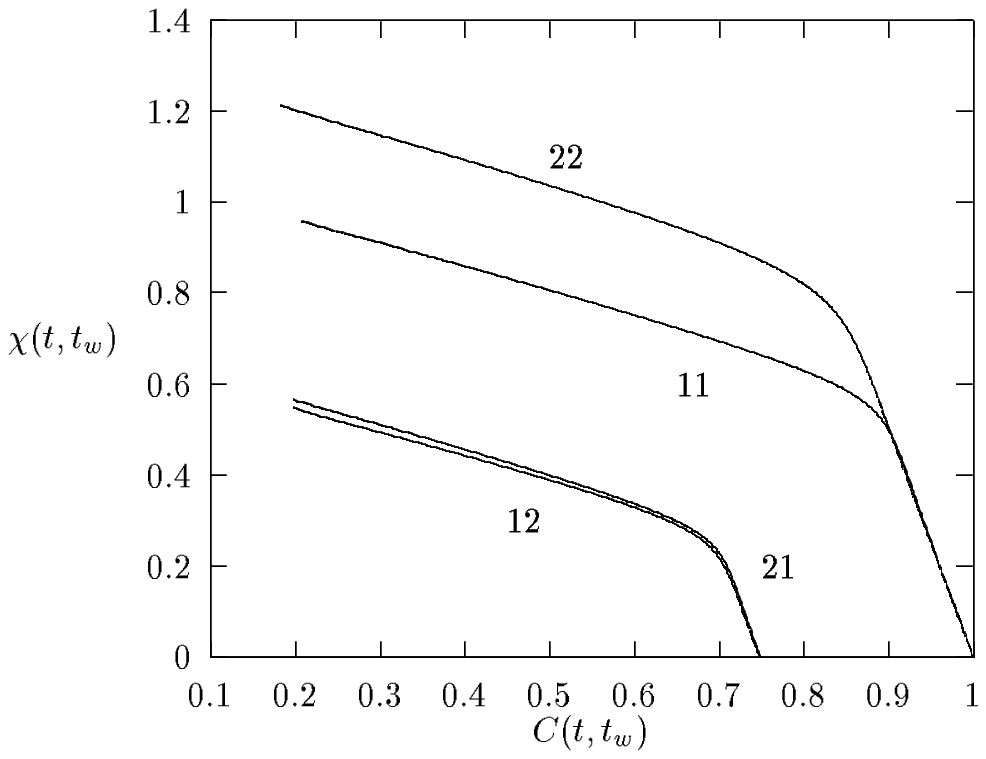}
\label{FIGtwosysth05}
}
\caption{The susceptibilities $\chi_{ij}(t,t_w)$ vs.\ 
the correlation functions $C_{ij}(t,t_w)$ for the two
  {\it weakly\/} coupled subsystems.
 All the curves become parallel: also the aging regimes have {\it thermalized\/}.
}
\end{figure}

The effective temperature indeed regulates thermalization, 
and we believe this the reason why it deserves its name.
The out of equilibrium dynamics of the model can then be interpreted as having
fast scales that are thermalized with the bath, and slow (aging) scales 
that are at a higher effective temperature. As time passes, more modes 
thermalize with the bath, and the frequencies that are in the aging
regime become lower and lower.
The effective temperature 
shares some, but not all, properties with the `fictive temperature'
\cite{To} of glass phenomenology (see Ref. \citen{cukupe} for a discussion).

\end{document}